\documentclass[]{aa}  
\usepackage{graphicx}
\usepackage{natbib}

\usepackage{txfonts}
\begin{document}

\title{The triple system HIP\,96515: a low-mass eclipsing binary with a  DB white dwarf companion\thanks{Based on observations 
collected at the Paranal Observatory under programs 77.C-0483(A) and 81.C-0826(A).}}
   \author{N. Hu\'elamo\inst{1}
          \and 
          L.P.R. Vaz\inst{2}
          \and
         C. A. O. Torres\inst{3}
         \and
          P. Bergeron\inst{4}
          \and
          C. H. F. Melo\inst{5}
          \and
          G. R. Quast\inst{3}
          \and
          D. Barrado y Navascu\'es\inst{1}
          \and
          M. F. Sterzik\inst{6}       
          \and
          G. Chauvin\inst{7} 
          \and
          H. Bouy\inst{8}\thanks{Marie Curie Outgoing International Fellow}
          \and
          N. R. Landin\inst{2} 
             }
\offprints{N. Hu\'elamo}
\institute{LAEX-CAB (INTA-CSIC); Postal address: LAEFF, P.O. Box 78, E-28691 Villanueva de la Ca\~nada,   Madrid, Spain\\
   \email{nhuelamo@laeff.inta.es}
          \and
          Depto.\, de F\'{\i}sica,
          Universidade Federal de Minas Gerais, C.P.702, 30161-970 --
          Belo Horizonte, MG, Brazil
         \and
         Laborat\'orio Nacional de Astrof\'{\i}sica/MCT, Rua Estados 
         Unidos 154, 37504-364 Itajub\'a, Brazil
         \and
          D\'epartement de Physique, Universit\'e de Montr\'eal, C.P.~6128,
          Succ.~Centre-Ville, Montr\'eal, Qu\'ebec H3C 3J7, Canada         
          \and
         European Southern Observatory, Karl-Schwarzschild-Strasse 2,
         D-85748 Garching bei Muenchen, Germany
         \and 
         European Southern Observatory, Alonso de Cordova 3107, 
         Casilla 19, Santiago, Chile
         \and 
         Laboratoire d'Astrophysique, Observatoire de Grenoble, BP 53, 
         38041 Grenoble, Cedex 9, France
         \and
         Instituto de Astrof\'{\i}sica de Canarias, C\/ 
         V\'{\i}a L\'actea, s/n, E38205 - La Laguna (Tenerife), Spain
}
   \date{Received; accepted}

 
  \abstract 
  {HIP\,96515\,A is a  double-lined spectroscopic binary  included in the SACY catalog
   as a potential young star, and classified as an eclipsing binary in the ASAS catalog. It has a visual
  companion (CCDM\,19371-5134\,B, HIP\,96515\,B) at 8\farcs6.  
  If bound to the primary, the optical and infrared colors of this wide companion are
  consistent with those of a white dwarf.}  
  {The aim of this work is to characterize the system HIP\,96515\,A\&B 
  by studying each of its components.}  
  {We have analyzed spectroscopic and photometric observations of
  HIP\,96515\,A and its visual companion, HIP\,96515\,B.   To confirm the system 
  as a common proper motion pair, we have analyzed the astrometry of the components using 
  high-angular resolution infrared observations obtained within a time span of two years and archival astrometry. }
  {The high-resolution optical spectrum of HIP\,96515\,A has been used
  to derive a mass ratio, $M_2/M_1$, close
  to 0.9, with the components showing spectral types of M1 and M2.
  The optical light-curve of HIP\,96515\,A shows periodic variations
  with $P_{\rm orb}$=2.3456\,days, confirming that
  HIP\,96515\,A is an eclipsing binary, with preliminary parameters
  of $i$=$89\fdg 0$$\pm$$0\fdg 2$, and $M_1$=0.59$\pm$0.03\,M$_{\odot}$ and 
  $M_2$=0.54$\pm$0.03\,M$_{\odot}$, for the primary and secondary,
  respectively, at an estimated distance of 42$\pm$3\,pc.  
  This is a new eclipsing binary with component masses below 0.6\,M$_{\odot}$.
  
  Multi-epoch observations of HIP\,96515 A\&B show 
  that the system is a common proper motion pair.  The optical spectrum
  of HIP\,96515\,B is consistent with a pure helium atmosphere (DB) white
  dwarf. The comparison with evolutionary cooling sequence models 
  provides $T_{\mathrm{eff,WD}}$=19,126$\pm$195\,K, $\log g_{\mathrm{WD}}$=8.08,
  $M_{\mathrm{WD}}/\mathrm{M}_{\odot}$=0.6, and a distance of $\sim$46\,pc.  
  The estimated WD cooling age is $\sim$100\,Myr and the total age of the object 
  (including the main-sequence phase) is $\sim$400\,Myr. Finally, if HIP\,96515 A\&B are coeval, and assuming a common age of $\sim$ 400\,Myr, the comparison of the masses of the eclipsing binary members with evolutionary tracks shows that they are underestimated by $\sim$15\% and    10\%, for the primary and secondary, respectively.
   }
  {} 
  \keywords{Stars: white dwarfs  --  Stars: binaries: eclipsing -- Stars: individual (HIP\,96515)}

 \authorrunning {Hu\'elamo et al.}
 \titlerunning {The triple system HIP\,96515}
    \maketitle
%

\section{Introduction}

In recent years, several infrared surveys have been focused on the
detection of substellar companions around young and nearby stars using high-contrast
imaging techniques \citep[e.g.][]{Chauvin2003,Lowrance2005, Masciadri2005,2007gdps}.
Young nearby stars are privileged targets for these kind of studies due to their proximity
(below 100 pc) which allows to inspect very small separations, and 
their youth, which enhances the contrast between the central star and the potential companion.

In 2006 we started an  infrared survey to detect substellar
objects around young stars from the SACY catalog \citep{Torres2003a,Torres2006}.
Our sample included HIP\,96515, a nearby \citep[d=43.9$\pm$9 pc, ][]{Leeuwen2007} M1-type star
classified as a double-line spectroscopic binary (SB2) by \citet{Torres2006}.
The object has a visual companion at an angular separation of 
$\rho $$\sim$ 8\farcs6 \citep[CR23, CCDM\,19371-5134\,B, ][HIP\,96515\,B hereafter]{WDSC1997, Domman2000}.
This visual binary has been detected in different  infrared and optical surveys, although its properties
have never been studied in detail.

As a first step to unveil the nature of the wide visual companion to HIP\,96515\,A,
we compared our {\em K$_\mathrm{s}$}-band photometry 
with evolutionary tracks \citep[][]{Baraffe2003, Chabrier2000}. 
The comparison suggested that HIP\,96515\,B may be of substellar
nature if the age of the primary is between 0.07-1 Gyr and the  two objects are co-moving.
However, the optical and infrared colors retrieved from archived observations are not
consistent with a substellar object but with a  white dwarf companion.

Motivated by this preliminary result, we decided to perform a detailed study of HIP\,96515.
In this work, we have analyzed new and archived observations of
the triple system and we have derived the main properties  of each source.
The paper is organized as follows: Section~2
describes the main observational properties of HIP\,96515 A\&B. In section~3, we analyze
spectroscopic and photometric data of the primary
star HIP\,96515\,A, while the true nature of the visual component, HIP\,96515\,B is unveiled
in section~4. A discussion of the system is provided in Section~5, and the main results 
are summarized in Section~6.

\begin{table}[t]
\caption{Main properties of  HIP\,96515\,A.}
 \label{stellar}
\advance\tabcolsep by -2pt
\begin{tabular}{ccccrr}
          \noalign{\smallskip} 
          \hline 
          \noalign{\smallskip}
 $\alpha$(J2000)&  $\delta$(J2000) &${\mathrm{Spectral}\atop{\mathrm{Type}}}$& ${\mathrm{Parallax^{1}}\atop{\mathrm{[mas]}}}$ & ${{\displaystyle{\mu_{\alpha}^1}}\atop{\mathrm{[mas~yr^{-1}]}}}$&${{\displaystyle{\mu_{\delta}^1}}\atop{\mathrm{[mas~yr^{-1}]}}}$ \\ 
          \noalign{\smallskip} 
          \hline 
          \noalign{\smallskip}       
$19^{\mathrm{h}}  37^{\mathrm{m}} 08^{\mathrm{s}}\!\!.7$ & $-51^{\circ} 34^{\prime} 00^{\prime\prime}\!\!\!.9$ &  M1$^2$ & 22.79 & 91.07 & -21.94    \\ [-3pt]  
&  &  &             $\pm$4.87&$\pm$4.93& $\pm$4.62   \\ 
\hline\hline
\end{tabular}

Notes: $^1$~\citet{Leeuwen2007}; $^2$~\citet{Torres2006}
\end{table}

\begin{table}[t]
\caption{Photometric data of the visual binary HIP\,96515\,A\&B}\label{phot}
\advance\tabcolsep by -4pt
\begin{tabular}{llcccl}
          \noalign{\smallskip} 
          \hline 
          \noalign{\smallskip}
 Target & ${{\displaystyle{V}}\atop{\mathrm{[mag]}}}$ & ${{\displaystyle{R}}\atop{\mathrm{[mag]}}}$ & ${{\displaystyle{J}}\atop{\mathrm{[mag]}}}$ & ${{\displaystyle{H}}\atop{\mathrm{[mag]}}}$  & ${{\displaystyle{K_{\mathrm{s}}}\atop{\mathrm{[mag]}}}}$   \\
 \noalign{\smallskip} 
          \hline 
          \noalign{\smallskip}
HIP\,96515\,A  & 11.78$^1$ & 10.90$^1$ &   8.82$^2$           &  8.15$^2$           &  \hphantom{$\pm$}7.96$^2$  \\[-3pt]
               &   $\pm$0.04\hphantom{$^2$~~\,}           & $\pm$0.03\hphantom{$^2$~~\,}& $\pm$0.03\hphantom{$^2$~~\,}&$\pm$0.04\hphantom{$^3$~~\,}&$\pm$0.03\hphantom{$^2$} \\
HIP\,96515\,B  & 13.0$^3$ &  14.3$^4$           &  14.54$^2$           &  14.44$^2$           &  14.52$^2$, 14.94$^5$ \\[-3pt]
                           &          &$\pm$0.3\hphantom{$^4$}&$\pm$0.18\hphantom{$^4$}&$\pm$0.51\hphantom{$^4$}& $\pm$0.31\hphantom{$^4$}$\hphantom{,}\pm$0.08\hphantom{$^4$}\\
\hline
\hline
\end{tabular}

Notes: $^1$~\citet{Torres2006}; $^2$~2MASS catalog \citep{2MASS}; $^3$~Visual Double stars in Hipparcos catalog \citep{Domman2000}; $^4$~UCAC2 catalog; $^5$~NACO/VLT data (this work).
\end{table}

\section{Observational properties of HIP\,96515 A\&B}

The main stellar properties of HIP\,96515\,A are
displayed in Table~\ref{stellar}, while its optical and infrared
photometry are shown in Table~\ref{phot}.

The analysis of  the optical spectrum of HIP\,96515\,A obtained during the SACY survey
has allowed \citet{Torres2006} to classify it as an SB2, with almost equal-mass
components (see Table~\ref{specdata}). From the cross-correlation function (CFF), 
we have estimated a magnitude difference of 
$\Delta\,V$=\,0.55\,mag between the binary components.

HIP\,96515 is included in the $ROSAT$ All-Sky Bright Star catalog \citep{voges1999}.
The source was detected by $ROSAT$ with a count rate of 0.172$\pm$0.036 counts\,s$^{-1}$.
We have derived the X-ray flux using the conversion factor definition by 
\citet{Schmitt1995}.
The X-ray luminosity, assuming a distance of 44 pc (see Table~1), 
is $L_{\mathrm{x}}$=(3.61$\pm$0.75)\,10$^{29}$erg\,s$^{-1}$.  Using the
bolometric correction ($BC$) by \citet{Kenyon1995} for an M1-type star, and the $V$-mag listed
in Table~\ref{phot},  we derive  a $\log(L_\mathrm{x}/L_\mathrm{bol})$ ratio of --3.07.  
If we assume that the two components of the spectroscopic binary (HIP\,96515\,Aa \& Ab) are emitting in 
X-rays, so that $L_{x,Aa}=L_{x,Ab}=L_{x}/2$, and using the $V$-mag of each of the
binary members (see Table~3) and the $BC$ for a M1 and M2 star, respectively, we derive 
$\log(L_\mathrm{x}/L_\mathrm{bol})$ ratios of --3.17 and --3.04 for Aa and Ab.
These values are comparable to those found in very active pre-main sequence (PMS)
late-type stars \citep[e.g.][]{barrado1999,stelzer2001} and in main-sequence (MS) late-type stars in
very  close binaries \citep[see e.g.][]{LopezM2007}.

The age of HIP\,96515\,A is not known.
The lithium content is normally used as a youth indicator in the case of late-type stars.   
However, it is not always present in young M-type stars  since 
it is depleted too fast \citep[e.g.][]{basri1996,neuh1997,Torres2008}.
In fact, none of the binary members of HIP\,96515\,A
show the lithium absorption doublet  ($\lambda$=6707.8~\AA) in the optical spectrum.
\citet{Torres2008} have studied the lithium content as a function of the {\em V-I} color
in different young associations with ages between 5-70 Myr. The 
comparison of HIP\,96515\,A with similar M-type stars (with {\em V-I}$\sim$1.8, \citet{Torres2006}) 
in those associations allows to compute a lower limit to the age of $\ge$ 70 Myr. By following the temporal evolution of lithium in 
stellar models \citep[eg. those by][]{landin06} 
for stars with the masses of HIP\,96515\,A (Sect.\,\ref{ebstudy}),
Li should be exhausted by an age between 30\,Myr (for the MLT parameter
$\alpha$=2) and 50\,Myr ($\alpha$=1).

Finally, HIP\,96515\,B, is included in the AC catalog \citep{Urban1998} as a visual
companion to HIP\,96515\,A.   The object has been detected in different infrared and optical 
surveys and multi-wavelength photometry is available in different public catalogs.
We have included all the available data in Table~\ref{phot}.  
Note that, in most cases, the photometric
measurements show large uncertainties.

\section{Characterization of HIP\,96515\,A: a M1+M2 eclipsing binary}
 \label{ebstudy} 

In order to characterize the SB2 in more detail, we retrieved
the photometric data of HIP\,96515\,A from both the Hipparcos archive
\citep[96 points]{HipCat} and from the "All Sky Automated Survey'',
ASAS\footnote{\small\tt http://www.astrouw.edu.pl/$\scriptstyle \sim$gp/asas/asas.html}
\citep[][ 818 points]{Asas2005}.  The Hipparcos observations, in its own photometric
system ($Hip$), were corrected with the
relation $V_{\rm J} = Hip - 0.08$, suggested for the $V-I$ of this
object \citep[$V-I$=1.85; ][]{Torres2006}.

\begin{figure}
   \includegraphics[width=0.5\textwidth]{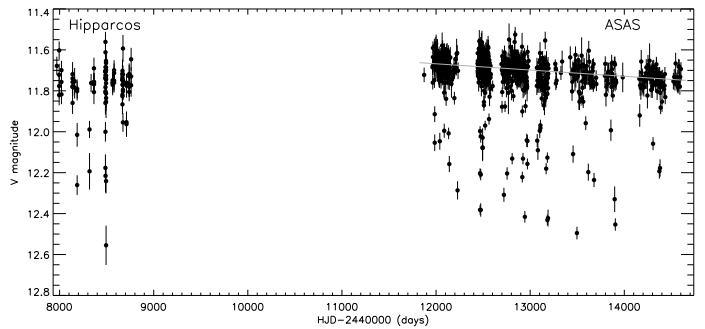}
   \caption{Hipparcos and ASAS V-band light-curves of HIP\,96515\,A}
   \label{lc}
\end{figure}

The observations of the source are displayed in
Figure~\ref{lc}, showing significant variability. 
The object has been classified as an eclipsing binary by \citet{AsasEB2006},
using the ASAS data alone. 
We have applied the method by \citet{lafler} 
to search for  any periodicity in the data. The best period, 
2.345 days, have been used to phase-fold both the ASAS and Hipparcos data, 
confirming the SB2 as an eclipsing system. 
In the next subsections, we  describe our full analysis of the photometric data and
derive the main properties of the eclipsing components, Aa and Ab.

\begin{table}\caption{Spectroscopic data of the eclipsing binary HIP\,96515\,A}
\begin{tabular}{llllll}
\hline
Star   &  {\em V}  & Spt. Type & v$_\mathrm{rad}$ &  H$\alpha$ EW & v.sin{\em i}\\ 
         &  [mag]     &                  & [km s$^{-1}$] & [\AA] & [km s$^{-1}$] \\ \hline
96515Aa & 12.29 & M1Ve & 84.1& 0.9 & 14.8$\pm$1.5 \\
96515Ab & 12.84 & M2:Ve& -81.3 & 2.0 & 15.8$\pm$1.6 \\ 
\hline 
\end{tabular}\label{specdata}
\end{table}

\subsection{Photometric data: Period search}
\label{period}

\begin{figure}
\includegraphics[angle=000,width=\hsize]{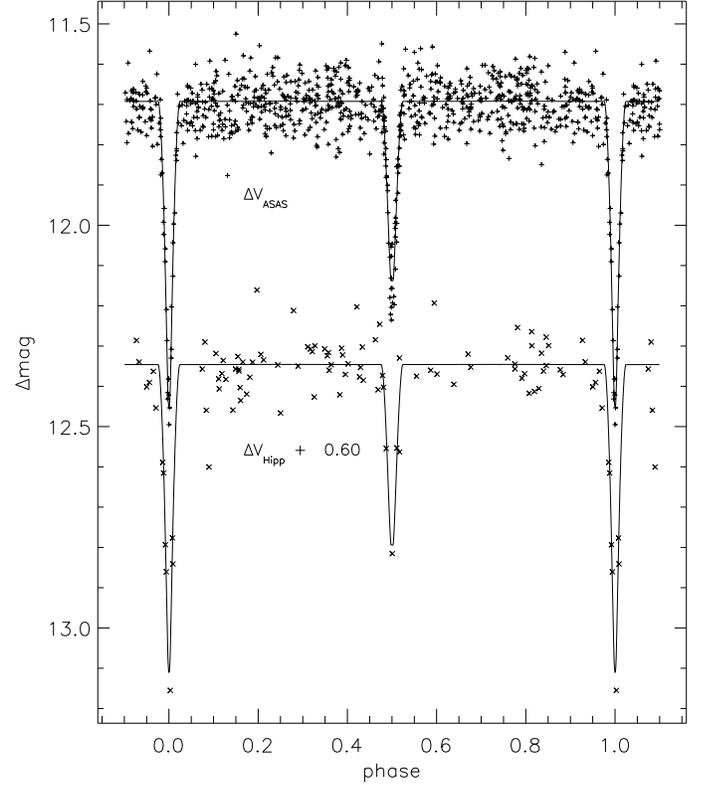}
\caption{The phase-folded light-curves of HIP~96515\,A from ASAS ($+$ symbols)
and Hipparcos ($\times$ symbols), consisting of 818 and 96 observations, respectively, with our theoretical
solution for $q$=0.9 obtained in Sect.\,\ref{wdanalysis}. The Hipparcos
observations were shifted by 0.6 mag, in order to separate the light curves.
Note that ASAS observations were not corrected from the long term variation
of the maxima, shown in Fig.\,\ref{lc}.
}
\label{Vlc}
\end{figure}

The method by \citet{lafler} was used to search for periods in the
Hipparcos V-band observations of HIP\,96515\,A. We searched for periods from
0.5 day to 63 days, by using a relative period step of $\delta
P/P$=4$\times$$10^{-7}$. Since M stars may flare, and HIP\,96515 is a
strong X-ray emitter, we excluded the 4 brightest measurements (from Hipparcos
measurements) in this
period search, in order to avoid these clearly non-periodic events, if
they are present.  Around 60 significant periods were found, but the
most significant was $P_{\rm orb}$=(2.344451$\pm$0.000002)\,d.  
In the analysis of the photometry with a version of the WD model (see
Sect.\,\ref{wdanalysis}), we adjusted a phase shift to the folded
observations, and obtained the ephemeris (HJD; the digits in parentheses
affect the last digits)
\begin{equation}
\label{eq:ephemeris}
{\rm Min~I} = 2,\!448,\!493.80515(10) + 2.3444507(15) E,
\end{equation}
which we use in our analysis. According to Eq,\,(\ref{eq:ephemeris})
the secondary minimum occurs very close to phase 0.5, indicating an
orbit with a small (if any) eccentricity, and in the lack of further evidence,
we assume in the preliminary analysis, a circular orbit. 
We note that circular orbits are consistent with short orbital (and rotational) periods, like
the one found in HIP\,96515\,A \citep[e.g.][]{Zahn1989}. 
The V light-curves 
(Hipparcos and ASAS), folded with these ephemeris, are shown in Fig.~\ref{Vlc}.
The stellar eclipses are clearly seen in both datasets.

\subsection{Preliminary analysis of the HIP\,96515\,A light-curve}
\label{wdanalysis}

We have analyzed the Hipparcos optical light-curve with a version
of the WD model \citep{wd1971,wilson1979,wilson1993} extensively
improved as described in \citet{uoph} and references therein. 
Our version uses stellar atmospheres and models the
radiated flux of both components using the PHOENIX atmosphere models
\citep{allardhauschildt1995,allardetal1997,hauschildtetal1997a,
hauschildtetal1997b}. The (linear-law) limb-darkening coefficients for
both components have been taken from \citet{claret2000}, and
interpolated using a bi-linear scheme for the current values of $\log
g$ and $T_{\rm eff}$ at each iteration.  The reflection albedos for
both components have been fixed to a value of 0.5, which is 
appropriate for stars with convective envelopes. 
The gravity-brightening exponent,
$\beta$, has been computed using the local value of $T_{\rm eff}$ for each
point on the stellar surface and taking into account mutual illumination 
following \citet{alencarvaz1997} and \citet{alencaretal1999}.  Our
procedure combines the WD modified code with several {\small\tt UNIX}
scripts and auxiliary programs to guarantee the consistency 
of the solutions in all steps.

We have a single measurement of the radial velocity for each component
at the phase 0.7715, according to Eq,\,(\ref{eq:ephemeris}), by chance,
very close to the quadrature (0.75 in a circular orbit), when the
line separation is maximum. Since we
do not have any estimation of the center of mass velocity of the
system, the mass ratio is undefined. Therefore, to investigate
the solution space, we have decided to
make a grid of solutions for different values of the mass ratio,
$q$(=$M_2$/$M_1$), varying from 1.000 to 0.850 in steps of 0.025.
By making this grid we can determine
the velocity of the center of mass of the system for each $q$, and
adjust the rest of parameters to reproduce the individual
heliocentric velocities measured at phase 0.7715. 

\begin{table}[t]
\caption
{Grid of solutions for 7 values of the mass ratio, $q$(=$M_2$/$M_1$), for HIP\,96515\,A.
Note that $V_1$ and $V_2$, below are the theoretical velocities for the
primary and the secondary at the orbital phase $\phi$=0.7715.
All the solutions have the reduced $\chi^2$ the closest to 1 possible, for both data sets.
}
\label{grid}
\centering
{\scriptsize
\advance\tabcolsep by -3.0pt
\begin{tabular}{lrrrrrrr}
\hline \hline
\noalign{\vskip 1pt}
$q$ (=$M_2$/$M_1$)& 1.000 & 0.975 & 0.950 & 0.925 & 0.900 & 0.875 & 0.850 \cr
\hline
\noalign{\vskip 2pt}
$V_1$ (km s$^{-1}$)& 82.70 & 81.65 & 80.60 & 79.48 & 78.35 & 77.19 & 75.99 \cr
$V_2$ (km s$^{-1}$)& $-$82.70 & $-$83.75 & $-$84.82 & $-$85.92 & $-$87.05 & $-$88.21 & $-$89.40 \cr
$V_{\gamma}$ (km s$^{-1}$)& 1.40 &  2.45 & 3.52 & 4.62 & 5.75 &6.91 & 8.10 \cr
$K_{1}$ (km s$^{-1}$)& 83.46 & 82.41 & 81.32 & 80.21 & 79.07 & 77.90 & 76.69 \cr
$K_{2}$ (km s$^{-1}$)& 83.46 & 84.51 & 85.60 & 86.71 & 87.85 & 89.02 & 90.23 \cr
$M_{1}$ ($M_{\odot}$)& 0.56 & 0.57 & 0.58 & 0.59 & 0.59 & 0.60 & 0.61 \cr
$R_{1}$ ($R_{\odot}$)& 0.64 & 0.65 & 0.65 & 0.64 & 0.64 & 0.64 & 0.64 \cr
$\log g_1$ (cgs)& 4.57 & 4.57 & 4.58 & 4.59 & 4.60 & 4.60 & 4.61 \cr
$M_{2}$ ($M_{\odot}$)& 0.56 & 0.56 & 0.55 & 0.54 & 0.53 & 0.53 & 0.52 \cr
$R_{2}$ ($R_{\odot}$)& 0.53 & 0.53 & 0.53 & 0.54 & 0.55 & 0.52 & 0.52 \cr
$\log g_2$ (cgs)& 4.74 & 4.74 & 4.73 & 4.71 & 4.69 & 4.72 & 4.72 \cr
$a$ ($R_{\odot}$)& 7.73 & 7.73 & 7.73 & 7.73 & 7.73 & 7.73 & 7.73 \cr
$\Delta m$ & 0.875 & 0.893 & 0.872 & 0.832 & 0.794 & 0.891 & 0.888 \cr
$i$ ($^{\circ}$) & 89.27 & 89.45 & 89.28 & 89.08 & 89.00 & 89.51 & 89.62 \cr
$d$ (pc)& 42 & 42 & 42 & 42 & 42 & 42 & 41 \cr
\hline
\end{tabular}
}
\end{table}

We have assigned an arbitrary initial value of $89\degr$
to the system inclination,  and we have used the
WD model to find initial values to the gravitational pseudo-potentials (which
define the size and form of the components) in order to reproduce the optical
light-curve.  Finally, the initial values of the effective temperatures have been adopted from 
the tabulated values listed in \citet{Popper:80}, and $T_{\mathrm{ eff,pri}}$
was kept fixed at the adopted value of 3,714$\pm$150\,K. In the lack of
further information, the eclipsing components were considered to be
synchronized with the orbital motion, an assumption that may not be true
(see below).

Despite the presence of a visual companion to HIP\,96515\,A at
$8\farcs6$, we do not expect any contamination in the Hipparcos light-curve of the 
primary, since HIP\,96515\,B is outside the Hipparcos field of
view. On the other hand, the resolution of the ASAS setup is 14\farcs2/pixel
and the light from HD\,96515\,B is definitely included in their measurements.
Therefore, we will assume a small amount of third light contribution in
the analysis of ASAS light curve.
From the appearance of the folded light curve (using our
ephemeris) and, in the lack of a well determined radial velocity curve,
we have assumed a circular orbit for our analysis.

Taking into account the above mentioned initial values and
assumptions, we have modeled the orbital inclination, the
gravitational pseudo-potentials, the secondary effective temperature, a phase shift of the
primary minimum and the primary internal luminosity for each point of
the grid.  The most interesting results from our modeling are
displayed in Table \ref{grid}.

\begin{figure}[t]
\includegraphics[angle=000,width=\hsize]{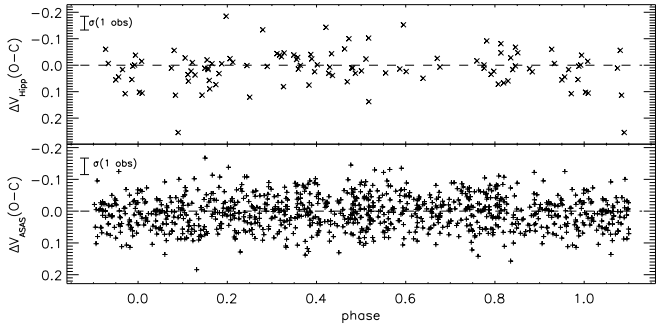}
\caption{The residuals O$-$C between the $V$ light curves of HIP~96515\,A
from Hipparcos ($\times$ symbols) and ASAS ($+$ symbols) shown in
Fig.\,\ref{Vlc}, and our theoretical solution for $q$=0.9 obtained in Sect.\,\ref{wdanalysis}. The standard
error (1 observation) is shown in the upper left part of each panel.}
\label{Vo-c}
\end{figure}

After convergence of the solutions of Table \ref{grid}, we adjusted
the amount of third light of the $q$=0.900 solution, due to the evidence
that the visual companion, HIP\,96515\,B,  contaminates the ASAS measurements. The third light
parameter, in units of the eclipsing system light at quadrature, determined
was ${\cal L}_3$=0.005$\pm$0.038 and the fit was
essentially indistinguishable from the solution with ${\cal L}_3$=0.

In order to reproduce the observed radial velocities, we have adjusted
the semi-major orbital axis which, together with the orbital
inclination and period, fixed the theoretical radial velocity amplitudes, $K_1$
and $K_2$  (in km s$^{-1}$). The apparent V-magnitude retrieved from SIMBAD is
$11.10$\,mag. However, the Hipparcos measurement converted to Johnson`s V
gives $11.73$\,mag and our measurement is $11.78$\,mag \citep{Torres2006},
value adopted for the distance estimation.
By using bolometric corrections for both components from
\citet{Popper:80}, we can derive the absolute dimensions and, finally,
estimate the distance to be (42$\pm$3)pc. The interstellar
absorption in such a short distance should be negligible. The parallax
for the system is (22.79$\pm$4.87)\,mas measured by Hipparcos, which
gives (44$\pm$9)\,pc, in very good agreement with our determination.

All solutions of Table\,\ref{grid} are essentially indistinguishable
from the curve fitting point of view, as can be seen by the rms of the
residuals ($\sigma$, 1 obs) in magnitudes. The probable intrinsic
variability of one or both components makes it necessary to follow the
system photometrically closely in time, preferably in different
passbands and simultaneously with the spectroscopy for radial velocity
measurements.  Only with more precise light-curves in different
photometric bands and with a precise radial velocity curve it
will be possible to eliminate this degeneracy of the solutions in the
mass ratio.  For the purpose
of this paper, we will assume that the components are in the main-sequence 
and make use of the mass-luminosity relation to fix $q$.  

\begin{figure}[t]
\includegraphics[angle=000,width=\hsize]{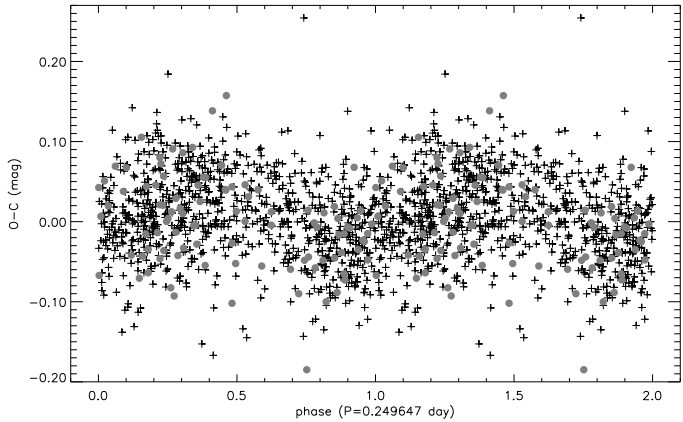}
\caption{The residuals O$-$C between the $V$ light curves of HIP~96515\,A
from ASAS ($+$ symbols) and Hipparcos (grey circles) shown in
Fig.\,\ref{Vo-c},  folded with the period of (0.24965$\pm$0.00001) day.
Note that the pattern is present in both Hipparcos and ASAS data sets
(two cycles are plotted).
}
\label{omcvsper}
\end{figure}

Figure \ref{Vlc} shows our solution for $q$=0.9 and Fig.\,\ref{Vo-c} shows
the corresponding residuals O$-$C. The steady and slow decreasing of the
maxima present in ASAS observations was not corrected for in this analysis.
It is common that M dwarfs present variations \citep[e.g.][]{2006A&A...448.1111R}
mainly due to spots on their surfaces. However, we cannot say, with the
data we analyzed, if the variability is due to one or to both eclipsing
components. We searched for periodicities in the O$-$C residuals from
0.08 day to 50 days, using the phase dispersion method by \citet{lafler}
and relative period steps of $\delta P/P$=3.d-3, and
found a significant period of (0.24965$\pm$0.00001) day (about 6 hrs), as can be
seen in Fig.\,\ref{omcvsper}. It is interesting to note that the variation
pattern is present in both the Hipparcos and the ASAS data sets.
The cadence of these observations (one or two observations a day, being common large
intervals without observations) is not adequate for detecting these
systematic variations in the O$-$C, with the annoying consequence of
introducing a multitude of aliases in the analysis. It is, however,
clear that the period of 0.24965 day, if real, is not the only one:
Fig.\,\ref{lc} gives indication for longer period variations and we, as
a matter of fact, estimate at least another possible period in the
O$-$C of (4,100$\pm$200) days, but with little significance level, due
to the relatively short interval of the observations ($\sim$6,630 days).

The period of 0.24965\,day detected may be due either to pulsation or to
spots on the surface of one or both of the eclipsing components. 
\citet[][]{2006A&A...448.1111R} list other M dwarfs with 6 hrs period.
However, if the variation is due to spots on HIP\,96515\,A, it would indicate
that the spotted component(s) spins significantly (more than 9 times)
faster that the orbital motion \citep[see e.g. CV~Boo][]{TG2008}.
This is probably not the case in HIP\,96515\,A: 
using an orbital period of 2.345\,days, and
the radius of the binary components from Table~5, we estimate rotational velocities of $\sim$14\,km.s$^{-1}$
and $\sim$12\,km.s$^{-1}$ for Aa and Ab, respectively. These values are comparable with the
ones derived from the optical spectrum (Table~3, and assuming an inclination of 89 degrees), 
indicating that the system is synchronized. Therefore, the origin of the 6h-period 
remains unclear with the current data-set.

\begin{table}[h]
\setlength{\tabcolsep}{0.05\tabcolsep}
\caption[]{Preliminary physical parameters for HIP~96515\,Aa\&Ab
}
\label{absdim}
\begin{flushleft}
\begin{tabular}{lrclrcl}
\noalign{\hrule\smallskip}
 & \multicolumn{3}{c}{Primary} & \multicolumn{3}{c}{Secondary} \\
\noalign{\smallskip\hrule\smallskip}
\multicolumn{5}{l}{Absolute dimension:} \\
Mass (M$_{\odot}$) & ~~~ 0.59 &$\pm$& 0.03~~~ &~~~ 0.54 &$\pm$& 0.03~~~ \\
Radius (R$_{\odot}$) & 0.64 &$\pm$& 0.01      &   0.55  &$\pm$& 0.03  \\
$\log g$ (c.g.s.)  &  4.59  &$\pm$& 0.03      &   4.69  &$\pm$& 0.05  \\
\multicolumn{5}{l}{Photometric data:} \\
$\log\,T_\mathrm{eff}$(K) & 3.571   &$\pm$& 0.018  &   3.555 &$\pm$&0.019  \\
$\log L/L_\odot$    &$-$1.15  &$\pm$& 0.07  & $-$1.35&$\pm$& 0.08  \\
$M_{\rm bol}$       &    7.63  &$\pm$& 0.18   &    8.13  &$\pm$& 0.20   \\
$M_{\rm V}$         &    9.10  &$\pm$& 0.18   &    9.90  &$\pm$& 0.20   \\
$L_{\rm sec}/L_{\rm pri}$ &\multicolumn{6}{c}{0.48$\pm$0.24}              \\
P$_{orb}$ (days) & \multicolumn{6}{c}{2.3456\phantom{10}} \\
$i$ ($\deg$) & \multicolumn{6}{c}{89.0$\pm$0.2\phantom{10}} \\
Distance (pc)         &\multicolumn{6}{c}{42$\pm$3\phantom{10}}         \\
\noalign{\smallskip}\hline
\end{tabular}
\end{flushleft}
\end{table}

By combining the results of Table\,\ref{grid}, and taking into account the
maximum variations of the variables along the grid, we can derive the
preliminary absolute dimensions for HIP~96515\,A, which are 
listed in Table\,\ref{absdim}. The masses of the binary components,
HIP\,96515 Aa and Ab, are below  0.6 M$_{\odot}$. Hence, HIP\,96515\,A
is the fourteenth confirmed EB with component masses below 0.7 M$_{\odot}$
\citep[see e.g. ][]{Shkolnik2008}, which are important to calibrate theoretical 
evolutionary models.

Finally, we note that the eclipsing system definitely deserves more observations, both multi-band photometry 
with a cadence that allows determination of
precise light curves, and spectroscopy, in order to obtain 
reliable radial velocity curves and spectral types of both components. If possible, 
HIP\,96515\,B, should also be observed simultaneously, in order to give consistency
to the analysis.

\section{Characterization of HIP\,96515\,B}

In the next subsections, we will focus on the characterization of HIP\,96515\,B.
First, we have investigated if the object forms a common proper
motion pair with HIP\,96515\,A. 
To do this, we have analyzed  diffraction-limited infrared observations of the pair
obtained in the course of our program to detect
substellar object around  SACY targets (Huelamo et al. 2008, in prep.).
As a second step, we have obtained an optical spectrum to derive the main properties
of the object.

\subsection{NACO/VLT observations of HIP\,96515}

We observed HIP\,96515 twice with NAOS-CONICA (NACO), the Adaptive Optics facility at the
Very Large Telescope (VLT), the nights of the 25th May 2006 and 16th June 2008. 
The target was observed with the visible wavefront sensor and the Ks-band filter. We used
the  S27 objective (nominal plate scale of $\sim$0.027\arcsec/pixel)  which  provides a
total  field-of-view of $\sim$27\arcsec$\times$27\arcsec. The total exposure time
was $\sim$12 minutes on source.

\begin{figure}[t!]
   \includegraphics[width=0.5\textwidth]{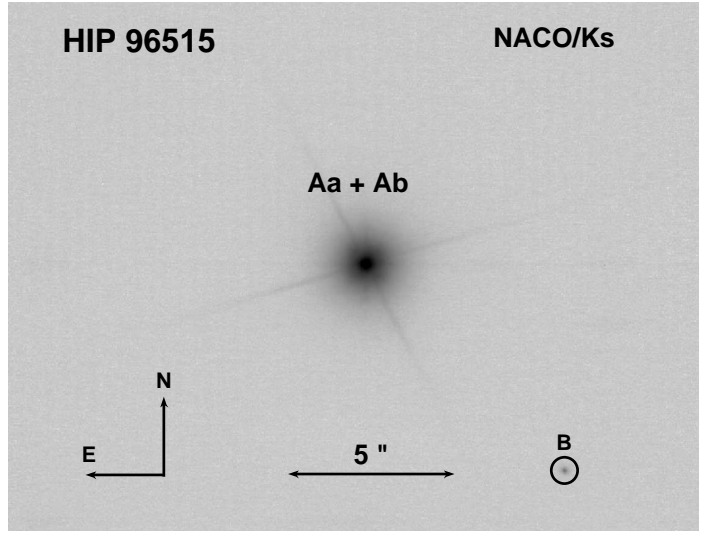}
   \caption{VLT/NACO {\em Ks}-band  image of HIP\,96515\,A and its visual companion, HIP\,96515\,B, obtained in June 2008. The projected separation is of 378\,AU at a distance of 44\,pc. Note that the eclipsing binary (Aa \& Ab) is not spatially resolved by NACO.}
   \label{nacoima}
\end{figure}

The data were obtained using a random jitter 
between the exposures to compute the sky emission. The images have been reduced
with the {\em Eclipse} reduction package \citep{Devillard1997}
following the standard steps: dark-subtraction, flat-field division 
and image shifting and stacking.
The final image of HIP\,96515 is displayed in Figure~\ref{nacoima}.
Apart from HIP\,96515\,A, we only detected the
 already known visual companion,  HIP\,96515\,B, at a projected
 separation of 8\farcs6  ($\sim$378\,AU at 44\,pc)  from the primary.
 Note that the eclipsing binary is not spatially resolved in the NACO image.

The difference in magnitude between HIP\,96515\,A\&B is 6.98$\pm$0.07\,mag. Taking into account
the brightness of the primary ($K_{\mathrm s }$=7.9\,mag, see Table~2), 
we can estimate  a $K_{\mathrm s }$-mag of 14.94$\pm$0.08 for the secondary.

The NACO images have been used to derive 
the separation and position angle of the two objects in two different epochs.
To derive accurate astrometry of HIP\,96515\,B in the two images, we have calibrated the
plate scale and orientation of the infrared detector, CONICA, with the astrometric calibrator
$\theta^1$ Ori~C \citep{mccaugh1994}. The results are displayed
in Table \ref{astromtab}.

\subsection{HIP96515 A \& B: A common proper motion pair}

The astrometric accuracy of the NACO/VLT data is high enough to  
investigate if HIP\,96515 A\&B are co-moving.
For completeness, we have  compared the NACO/VLT data  with older (and less accurate) 
astrometric measurements  from public surveys \citep[e.g.][]{Domman2000,2MASS}.
The visual binary is included in the Washington
Double Star Catalog \citep[WDSC,][]{WDSC1997} with three observations
between 1910 and 2000.  
Table \ref{astromtab} includes the angular separation and position angle
of the two components at five different epochs.
As seen, both remain almost constant within a time span of 98 years.

\begin{figure}[ht]
 \vspace{0.5cm}
   \includegraphics[width=0.42\textwidth]{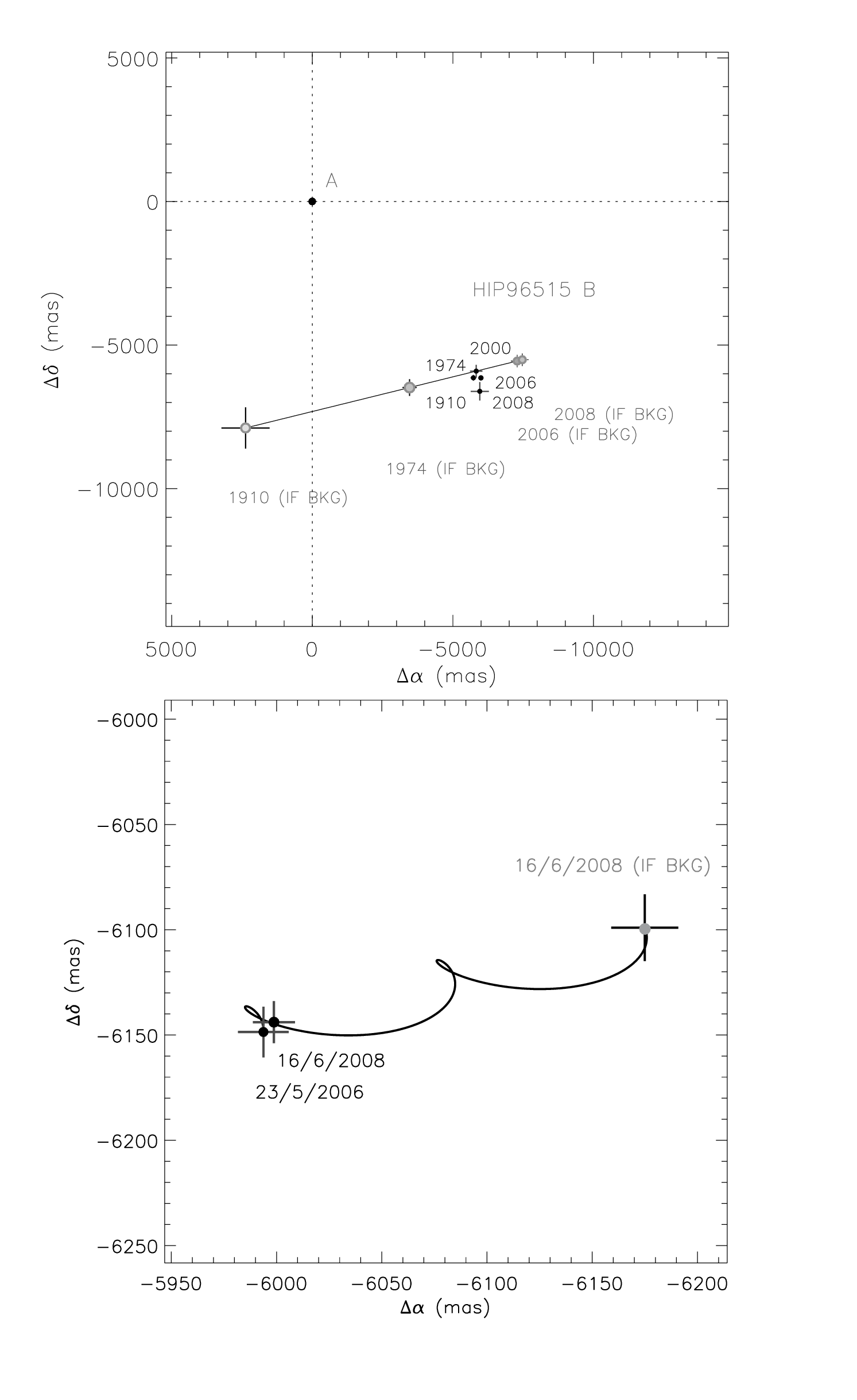}  
    \caption{Differences in right ascention and declination
    between HIP\,96515\,A \& B at different epochs (black circles). The expected position of a background object (grey circles and labels) at each epoch
    is also displayed. {\bf Top:} All the astrometric data from 1910 to  2008. {\bf Bottom:} Only the NACO data, which provides the best astrometric accuracy. The figure shows that the two objects are co-moving, that is, they are a common proper motion pair.}
   \label{astromfig}
   \end{figure}

\begin{table}
   \caption{Relative astrometry of HIP\,96515\,A \& B} 

  \advance\tabcolsep by -3.0pt  
  \begin{tabular}{lllll}
          \noalign{\smallskip} 
          \hline 
          \noalign{\smallskip} 
  Obs. Date & Separation       &  PA   & Reference\\
     & [arcsec]         & [deg] &        \\
     \noalign{\smallskip} 
          \hline 
          \noalign{\smallskip}
2008-06-16 & 8.586$\pm$0.008 & 224.3$\pm$0.2 & NACO/VLT, this paper\\
2006-05-23 &  8.586$\pm$0.010 & 224.2$\pm$0.2 & NACO/VLT, this paper \\ 
2000-10-10 &  8.3  $\pm$0.2    & 224.6 & 2MASS; \citet{2MASS} \\ 
1974.620      & 8.4          & 223  & \citet{Domman2000} \\ 
1910.705  & 8.9$\pm$0.3     & 222  & \citet{WDSC1997}\\ 
 \hline\hline
 \label{astromtab}
  \end{tabular}
 \end{table}

Figure~\ref{astromfig} (top panel) shows the difference in right
ascension (RA) and declination (DEC) of the binary members measured in
five epochs between 1910 and 2008. We have also overplotted the
expected difference in RA and DEC of a background object taking into
account the proper motion of the primary.  If we focus on the NACO/VLT
data only (Fig. ~\ref{astromfig}, bottom panel), which provides the
best astrometric accuracy, we can confirm that HIP\,96515 A\&B are
comoving , that is, they form a common proper motion pair.

\subsection{The Optical Spectrum of HIP\,96515\,B: observations and modeling}

To unveil the nature of HIP\,96515\,B, we have obtained an
optical spectrum of the target with the ESO Multi-Mode Instrument (EMMI) 
on the New Technology Telescope (NTT) on October 2007. 
The wavelength range goes from 380\,nm to 700\,nm, with a spectral
resolution of R$\sim$1100.  The spectrum has been reduced with
standard routines within IRAF, including bias subtraction and
flat-fielding, and has been flux-calibrated with a spectrophotometric
standard observed during the night.
\begin{figure}[t]
   \includegraphics[width=0.50\textwidth]{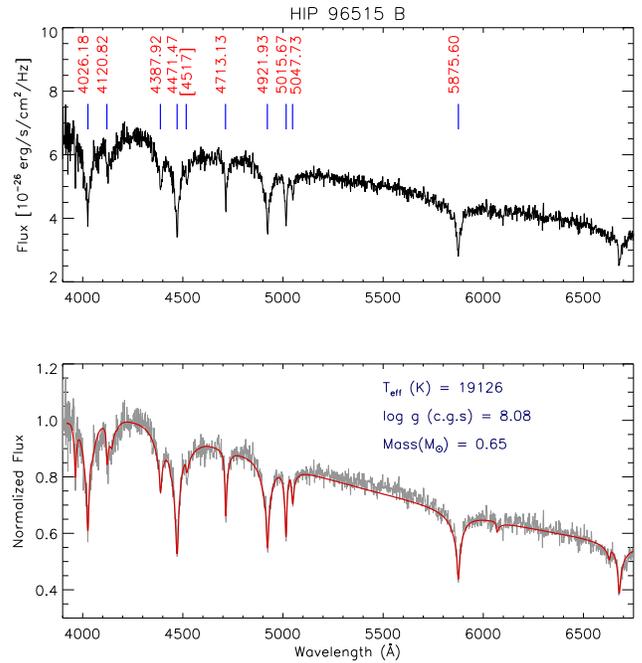}
   \caption{{\bf Top:} NTT/EMMI Optical spectrum of HIP\,96515\,B. 
    We have marked all the identified He I lines.     
    {\bf Bottom:} Fit of the optical spectrum using pure helium
   model atmospheres. 
   The best fit parameters are displayed at the top right corner. }
   \label{spec}
   \end{figure}
\begin{table}[h!]
\caption{Best fit parameters for HIP\,96515\,B}
\begin{tabular}{ll}
\noalign{\smallskip} 
          \hline 
          \noalign{\smallskip}
            $T_\mathrm{eff}$ (K) & 19126$\pm$195\\
  log {\em g}  (cgs)                      & 8.08$\pm$0.07 \\
  Mass (M$_\odot$) &  0.648\\
  Radius (R$_\odot$) & 0.01212\\
  $M_V$  (mag) & 10.89\\
  $M_J$  (mag)  &  11.33\\
  distance (pc)  &  46\\
  Progenitor Mass (M$_\odot$) & 3.1$\pm$0.8\\ 
  WD Cooling age (Myr)  & 106 $\pm$ 13\\
  MS lifetime (Myr)          & $\sim$ 300 $^1$\\    
  Total age (Myr)             & $\sim$ 400  \\  \hline
  \end{tabular}\label{modelDB}
 
 $^1$ see text
 
 \end{table}

The optical spectrum of HIP\,96515\,B is displayed in
Figure~\ref{spec}.  The continuum increases towards bluer wavelengths and
the main detected spectral features are the He\,{\sc I} absorption
lines, typical of helium (DB) white dwarfs. 

We have compared the optical spectrum of HIP\,96515\,B with
atmospheric models of DB white dwarfs, in order to derive the physical
parameters the object.
Our grid of model atmospheres and synthetic spectra for DB stars is
described in \citet{beauchamp96}. These include the improved Stark
profiles of neutral helium of \citet{beauchamp97}. The model
atmospheres used here assume a pure helium composition and cover a
range of $T_{\rm eff}=10,000~(1000)~16,000~(2000)~30,000$~K and $\log
g=7.0~(0.5)~9.0$. Our fitting technique relies on the nonlinear
least-squares method of Levenberg-Marquardt \citep{press86}, which is
based on a steepest descent method. The model spectra (convolved with
a Gaussian instrumental profile) and the optical spectrum are first
normalized to a continuum set to unity. The calculation of $\chi^{2}$ is
then carried out in terms of these normalized line profiles only. We
adopt a pure helium composition based on the lack of any H$\alpha$
absorption feature; additional models with mixed helium and hydrogen
compositions allow us to set a limit of $N({\rm H})/N({\rm
He}) <10^{-5}$.

As usual, there is a cool and a hot solution on each side of the
temperature at which the line strengths reach their maximum, but the
hot solution near $T_{\rm eff}$=25,000~K can easily be ruled out based
on the slope of the observed energy distribution. Our final solution,
$T_{\rm eff}$=19,130$\pm$190~K and $\log g$=8.08$\pm$0.07, is displayed in
Figure~\ref{spec}. These atmospheric parameters can be converted
into a mass of {\em M}=0.65$\pm$0.04\,M$_\odot$ using evolutionary models with
C/O cores, $q({\rm He})\equiv \log M_{\rm He}/M_{\star}$=$10^{-2}$, and
$q({\rm H})$=$10^{-10}$, which are representative of helium-atmosphere
white dwarfs\footnote{see http://www.astro.umontreal.ca/$\scriptstyle\sim$bergeron/CoolingModels}.

By using the best fit parameters of the model we can derive the
absolute magnitudes in different photometric bands (see
Table~\ref{modelDB}). Using $M_\mathrm{J}$ and the 2MASS {\em J}-band brightness of
HIP\,96515\,B, which is the measurement with the smallest photometric error,
 we estimate a the distance to the source of $d$=45.8\,pc,  which 
 is in good agreement with both the Hipparcos parallax and the distance 
 derived from the analysis of  HIP\,96515\,A.

The evolutionary sequences described above, as well as empirical relations \citep[see e.g.][]{Fontaine2001}, 
allow to estimate the mass of the white dwarf progenitor and its  cooling age. 
These data are provided in Table~\ref{modelDB}. 
We obtain two different estimates of the mass of the white dwarf
progenitor. First, we use the empirical initial-final mass relationship
(IFMR) from \citet{wood1992}, which is based on the study of
the white dwarf luminosity function and mass distribution \citep[see also][]{Leggett1998, Fontaine2001}. This
yields a value of $M_{\rm MS}=3.8 \pm 0.5$ M$_{\odot}$. Alternatively, we use
the IFMR determined by \citet{Ferrario2005}  based
on white dwarfs observed in open clusters (see in particular their
Fig. 1). This second relation yields a significantly smaller final
mass of $M_{\rm MS}=2.5 \pm 0.5$ M$_{\odot}$. However, as discussed by
Ferrario et al., the IMFR of most clusters appears to have an
intrinsic spread (i.e, there is a range of initial masses that gives
rise to a given final mass). Hence a value of $M_{\rm MS}$ as high as
3.8 M$_{\odot}$ cannot be necessarily ruled out. In the following, we thus
adopt a mean value of $M_{\rm MS}=3.1 \pm 0.8$ M$_{\odot}$ to allow for the
possible mass range of the progenitor.

We have derived the main-sequence lifetime of HIP\,96515\,B using 
two sets of evolutionary tracks for  a 3.1$\pm$0.8\,M$_{\odot}$ star.
First, we have used the ATON 2.4 models, which are a modified version of the models by D'Antona \& Mazzitelli \citep{dm94,dm97}, and are fully described in~\citet{landin06}. According to these evolutionary models, 
a 3.1$\pm$0.8\,M$_{\odot}$ star spends $280^{+346}_{-125}$\,Myr on the main-sequence, i.e., from the zero-age main-sequence (ZAMS) to the terminal age main-sequence (TAMS).  
On the other hand, \citet{Claret2004} models (Z=0.02) predict a span of $\sim$317\,Myr for the MS time of a 3.1\,M$_{\odot}$.
Assuming a MS lifetime of $\sim$300\,Myr  and a cooling time of $\sim$100\,Myr (see  Table~\ref{modelDB}), the total age of the system would be $\sim$400\,Myr, being the errors dominated by the uncertainty in the mass.

\section{Discussion}

\begin{figure}[t]
\includegraphics[width=0.5\textwidth]{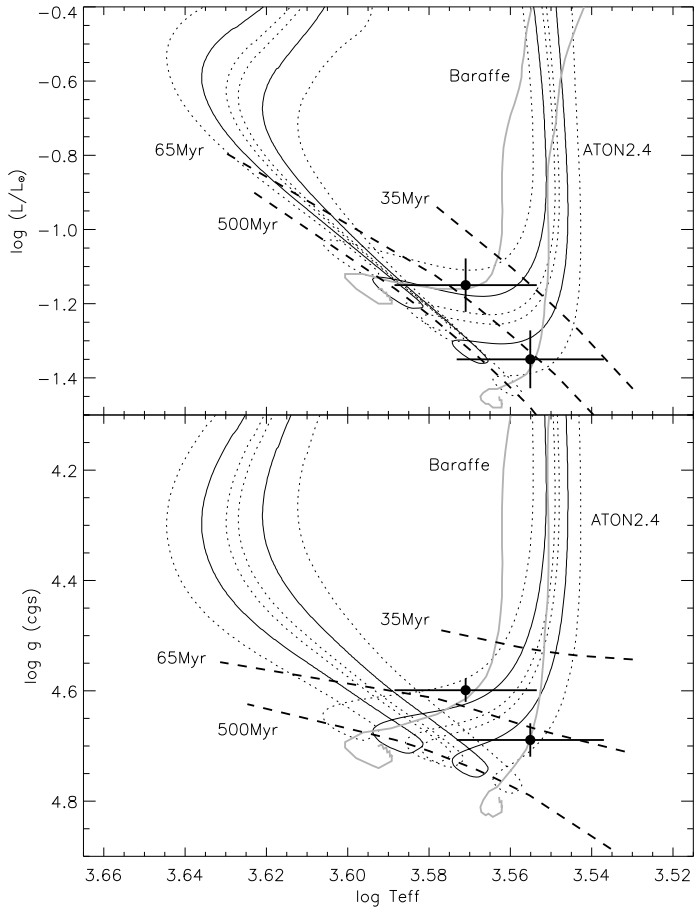}  
    \caption{Stellar Luminosity (top panel) and surface gravity (bottom panel) vs. effective temperature. 
    HIP\,96515\,Aa and Ab are represented by filled circles. The grey and black
    solid lines represent evolutionary tracks by~\citet{Baraffe1998} and~\citet{landin06}, respectively, for 0.6 and
    0.5\,M$_{\odot}$ stellar masses. The dotted lines  represent the uncertainties in the stellar masses for
    the ATON models. 
     The comparison of HIP\,96515\,A with the two sets of evolutionary tracks provides an 
     age of $\sim$60\,Myr, that is, places the eclipsing binary members on the pre-main sequence, although with large uncertainties.}
    \label{hrdiag}
   \end{figure}

The comparison of HIP\,96515 Aa \& Ab with evolutionary tracks by
\citet{Baraffe1998} provides an age of $\sim$50 Myr for
masses of 0.5\,M$_{\odot}$ (with alpha=1), and places the
binary on the PMS (see Fig.~\ref{hrdiag}).  
The PMS nature of HIP\,96515\,A is also suggested by tracks generated
with the ATON 2.4 stellar evolutionary code, where
we adopted the solar chemical composition (Z=0.0175 and Y=0.27) and 
alpha, $\alpha$, equal to 1. The tracks in Fig.~\ref{hrdiag} generated with ATON2.4 code
cover the PMS, MS and earlier stages of post main-sequence phases.
The best age, according to ATON 2.4, is
$\sim$60\,Myr (Fig.~\ref{hrdiag}), although with large uncertainties.

 On the other hand, our astrometric study shows that the white dwarf and 
the EB are co-moving. If we assume that HIP\,96515 A\&B
are coeval,  the age of the triple system should be $\sim$400\,Myr, 
which is significantly older than the age provided by the PMS tracks for HIP\,96515\,A,
although within the observational uncertainties of the primary star (see Fig.~\ref{hrdiag}).

To shed light on the evolutionary status of HIP\,96515\,A, we
have studied the kinematical properties of the object.
Using an average distance of 45\,pc (coming from the EB and WD independent studies presented here), the
RV derived by \citet{Torres2006}, and the proper motions measured by Hipparcos (see Table~1), we can derive
the {\em UVW} components of the  Galactic space velocity vector: {\em UVW}=$-6.0$, $-0.5$, $-21.6$\,km\,s$^{-1}$. We have compared them with the {\em UVW} vectors of young moving groups in the solar
neighborhood included in  \citet{Zuck2004} and \citet{Torres2008}. We conclude that HIP\,96515\,A
does not share the kinematical properties of  any of these associations.  

Assuming an age of 400\,Myr for the triple system, a possible explanation to the age of 60\,Myr derived with 
PMS evolutionary tracks  is the different evolution of EBs and very magnetically active stars 
in comparison with non-active objects. EBs are normally
fast rotators ($P_{\rm rot} <$\,3\,days) and they exhibit strong X-ray emission related with the presence of magnetic fields \citep{LopezM2007}.
As explained by e.g. \citet{Chabrier2007}, rapid rotation and/or magnetic fields reduce the efficiency 
of large-scale thermal convection in their interior leading to less efficient heat transport.
The reduction of the stars radiating surface due to the presence of cold spots can yield to smaller $T_{\rm eff}$ and
larger radii, thus departing significantly from the models predictions \citep[see e.g.][]{TorRib2002}.

\begin{figure}[t!]
\includegraphics[width=0.5\textwidth]{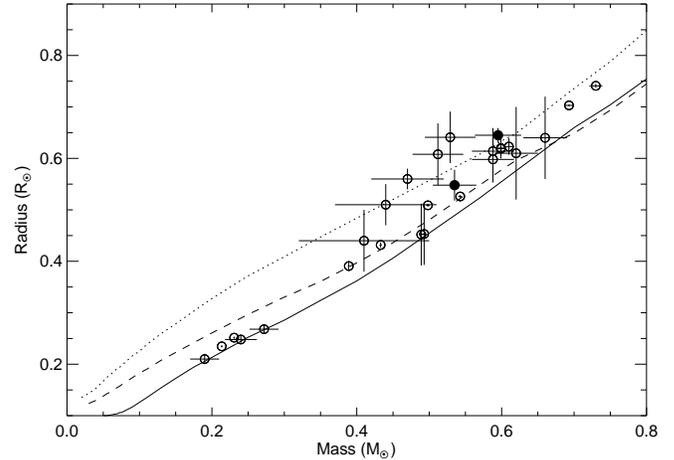}  
    \caption{Mass-Luminosity relation for all known eclipsing binaries with 
    masses $M<$ 0.7\,M$_{\odot}$ \citep[see][and references therein]{Shkolnik2008}. HIP\,96515 Aa and Ab are represented by filled circles. The dotted, 
    dashed and solid lines represent evolutionary tracks by~\citet{Baraffe1998} for 50, 100 and 500\,Myr, respectively.}
    \label{mlfig}
   \end{figure}

To investigate this in more detail, we have  studied the Radius vs. Mass relation for HIP\,96515\,A and compared it with the predictions from evolutionary models by \citet{Baraffe1998}.  The adopted models correspond to a solar metallicity of [Fe/H]=0, and a mixing length of $\alpha$=1. The results are displayed in Fig.~\ref{mlfig}, which
 includes all the low-mass eclipsing binaries confirmed so far. The data has been adopted from
\citet{Shkolnik2008} and references therein. 

 If we assume an age of 400\,Myr for the triple system, the radii of HIP\,96515\,Aa\&Ab are $\sim$\,15\% and 10\% larger, respectively,  than predicted by the evolutionary tracks (see Fig.~\ref{mlfig}).
 These discrepancies have been previously observed in similar eclipsing binaries 
 and have been attributed to the effect of rapid rotation, magnetic activity and/or 
 metallicity \citep[e.g.][]{Berger2006,Chabrier2007, LopezM2007}.

\section{Results and conclusions}

In this paper we have studied in detail HIP\,96515, a triple system composed by
a double-lined spectroscopic binary (SB2) and a co-moving white dwarf at a projected
separation of 8\farcs6.  We have analyzed both new and archived observations
of the system. Our main results can be summarized as follows:

\begin{itemize}

\item The analysis of multi-epoch optical photometry of HIP\,96515\,A
has revealed that this SB2 is an eclipsing binary with almost equal-mass
components: $M_{Aa}$=0.59\,M$_{\odot}$ and $M_{Ab}$=0.54\,M$_{\odot}$.  
In fact, this is the fourteenth confirmed eclipsing binary with
component masses below 0.7\,M$_{\odot}$, which are of
extreme importance to calibrate theoretical evolutionary tracks.

  \item Archival astrometry and new NACO/VLT data show that HIP\,96515 A\&B is a common proper motion pair.

\item The optical spectrum of HIP\,96515\,B is consistent with a hot
DB white dwarf. We have modeled it using evolutionary cooling
sequence models, deriving a temperature, surface gravity and mass of
$T_{\mathrm{eff}}$ (K)=19126, $\log g$ = 8.08, and M=0.65\,M$_\odot$.
Empirical relations  provide a progenitor mass of 3.1$\pm$0.8\,M$_{\odot}$, 
and a white dwarf  cooling age of 100$\pm$13\,Myr. 
The MS lifetime of a 3.1\,M$_{\odot}$ star, as derived from
evolutionary models, is $\sim$300\,Myr, resulting in a total age of $\sim$40\,Myr for the white dwarf.

 \item A direct comparison of the EB members with evolutionary tracks
  \citep[][]{Baraffe1998,landin06} places them on the PMS with an age of
  $\sim$60\,Myr, although with large uncertainties.   
  We note that HIP\,96515\,A does not share the the kinematical properties of  
  known young moving groups in the solar vicinity.

 \item If  HIP\,96515 A\&B are coeval, the system must be older than 300\,Myr.
  The discrepancy between this age and the 60\,Myr provided by PMS tracks  might be explained by the different
  evolution of  EBs, which normally show
 fast rotation and strong magnetic fields,  compared to non-magnetic stars. 
 Since HIP\,96515\,A shows fast rotation and strong X-ray emission, it  
 can depart  from the predictions of theoretical models, which place
 the binary on the PMS. 
 
\item If we assume an age of $\sim$400\,Myr for the triple system,  
the comparison of HIP\,96515 Aa\&Ab  with evolutionary tracks by \citet{Baraffe1998}  shows that the
models underestimate the stellar radii by 15\% and 10\%, for the primary and the secondary, respectively.
This behavior has been previously observed in similar low-mass EBs.

\end{itemize}

Finally, we note that additional RV observations are needed  to better constraint the orbit and
physical parameters of the eclipsing binary.

\begin{acknowledgements}
  We are very grateful to I. Saviane for obtaining the NTT/EMMI
  spectrum presented here.  NH is indebted to the Spanish {\em Programa
  Juan de la Cierva}. This research has been funded by Spanish grant MEC
  ESP2007-65475-C02-02 and MEC/Consolider-CSD2006-0070. 
  NH and CT gratefully acknowledge support from ESO-DGDF
  2007 program.
  GC acknowledges support from the Faculty of the European Space Astronomy Centre (ESAC).
  HB acknowledges the funding from the
  European Commission's Sixth Framework Program as a Marie Curie
  Outgoing International Fellow (MOIF-CT-2005-8389).  LPRV and NRL gratefully
  acknowledges financial support from the Brazilian agencies CAPES,
  CNPq and FAPEMIG.  PB is a Cottrell Scholar of the Research Corporation for Science 
  Advancement and is supported in part by the NSERC Canada and by the Fund 
  FQRNT(Qu\'ebec). This research has made use of the SIMBAD
  database, operated at CDS, Strasbourg, France, of NASA's
  Astrophysics Data System Abstract Service, and of the {\em
  Washington Double Star Catalog} maintained at the U.S. Naval
  Observatory. 

\end{acknowledgements}

\bibliographystyle{aa}
\bibliography{wd}

\end{document}